\documentclass[%
reprint,
 amsmath,amssymb,
 aps,
]{revtex4-1}

\usepackage{graphicx}% Include figure files
\usepackage{dcolumn}% Align table columns on decimal point
\usepackage{bm}% bold math
\usepackage{color}
\DeclareMathOperator{\atan}{atan}
\DeclareMathOperator{\csch}{csch}
\DeclareMathOperator{\sgn}{sgn}

\newcommand{\be}{\begin{equation}}
\newcommand{\ee}{\end{equation}}
\newcommand{\bea}{\begin{eqnarray}}
\newcommand{\eea}{\end{eqnarray}}

\begin{document}

%\preprint{APS/123-QED}

\title{Full consideration of acoustic phonon scatterings in two-dimensional Dirac materials}
\author{Khoe Van Nguyen$^{1,2,3}$}
\email{nvkhoe@gate.sinica.edu.tw}
\author{Yia-Chung Chang$^{1,4}$}
\email {yiachang@gate.sinica.edu.tw}
\address{$^{1}$ Research Center for Applied Sciences, Academia Sinica, Taipei 115, Taiwan}
\address{$^{2}$ Molecular Science and Technology, Taiwan International Graduate Program, Academia Sinica, Taipei 115, Taiwan}
\address{$^{3}$ Department of Physics, National Central University, Chungli, 320 Taiwan}
\address{$^{4}$ Department of Physics, National Cheng-Kung University, Tainan 701, Taiwan}

\date{\today}

%\keywords{Suggested keywords}

\begin{abstract}

The in-plane acoustic phonon scattering in graphene is solved by considering fully inelastic acoustic phonon scatterings in two-dimensional (2D) Dirac materials for large range of temperature ($T$) and chemical potential ($\mu$). Rigorous analytical solutions and symmetry properties of Fermionic and Bosonic functions are  obtained. We illustrate how doping alters the temperature dependence of acoustic phonon scattering rates. It is shown that the quasi-elastic and ansatz equations previously derived for acoustic phonon scatterings in graphene are limiting cases of the inelastic-scattering equations derived here. For heavily-doped graphene, we found that the high-$T$ behavior of resistivity is better described by $\rho(T, \mu) \propto T(1 - \zeta_a\mu^2/3(k_BT)^2)$ rather than a linear $T$ behavior, and in the low $T$ regime we found $\tau^{-1} \propto (k_BT)^4$ but with a different prefactor (i.e. $\sim$ 3 times smaller) in comparison with the existing quasi-elastic expressions.  Furthermore, we found a simple analytic  "semi-inelastic" expression of the form $\tau^{-1} \propto (k_BT)^4/(1+ c T^3)$ which matches nearly perfectly with the full inelastic results for any temperature up to 500 K and $\mu$ up to 1 eV. Our simple analytic results agree well with previous first-principles studies and available experimental data. Moreover, we obtain an analytical form for the acoustic gauge field $\beta_A = 3\beta\gamma_0/4\sqrt{2}$. Our analyses pave a way for investigating scatterings between electrons and other fundamental excitations with linear dispersion relation in 2D Dirac material-based heterostructures such as bogolon-mediated electron scattering in graphene-based hybrid Bose-Fermi systems.

\end{abstract}

\maketitle

%\section{Current issues of electron-phonon interaction in 2D Dirac materials}
\section{\label{sec:level1}Introduction}
%\section{\label{sec:level1}Introduction\protect}

Ions in a crystal lattice at a finite temperature ($T$) vibrate around their equilibrium positions; consequently, they produce quasiparticles named phonons - quantum states of lattice vibrations, which in turn cause electrical resistivity ($\rho$) by scattering off conducting charged carriers in the lattice \cite{Ziman1960,Ashcroft1976,Lundstron2000}. In these quantum processes, energy and momenta must be conserved \cite{Ziman1960,Ashcroft1976,Lundstron2000}. In general, electrical resistivity is proportional to the electron-acoustic phonon (EAP) scattering rates. Therefore, accurate calculations of acoustic phonon scatterings are very important in extracting various characteristic quantities of doped 2D Dirac materials such as electrical resistivity, effective deformation potential, carrier mobility, Bloch-Gr\"{u}neisen temperature, heat transfer rate, optical, remote interfacial and intra-ripple flexural  phonon scatterings from experimental data
\cite{Pietronero1980,Woods2000,Ando2006,Stauber2007,Vasko2007,Hwang2008,Bolotin2008,Chen2008,Mariani2008,Morozov2008,Efetov2010,Castro2010,Borysenko2010,
Perebeinos2010,Kozikov2010,Zou2010,Viljas2010,Mariani2010,Konar2010,Li2010,Dean2010,Min2011,Sarma2011,Cooper2012,Kaasbjerg2012,Ochoa2012,Park2014,Li2014,Sohier2014,Lucas2016,You2019}, in designing graphene-based hypersonic and acousto-electric devices and high-frequency spectrometers \cite{Spector1962,Pippard1963,Nunes2012,Zhao2013,Andersen2019},  As pointed out in Refs. \cite{Perebeinos2010,Sohier2014},  models with different angular dependencies result in different numerical prefactors for graphene's quasi-elastic scattering rates by in-plane acoustic phonons. However, they share a common formula in the high $T$ regime \cite{Pietronero1980,Woods2000,Stauber2007,Vasko2007,Hwang2008,Bolotin2008,Chen2008,Efetov2010,Castro2010,Borysenko2010,Perebeinos2010,
Viljas2010,Kozikov2010,Mariani2010,Zou2010,Dean2010,Min2011,Sarma2011,Kaasbjerg2012,Cooper2012,Ochoa2012,Sohier2014,Li2014}
\begin{equation}
\tau_{HT}^{-1}(\epsilon_{sk}) = \frac{J_a^2 |\epsilon_{sk}| k_BT}{4\rho_mv_{LA}^2\hbar^3v_F^2},\label{eq1}
\end{equation}
where $\epsilon_{sk}=s v_F \hbar k$ describes the band structure near the Dirac point with $s=+(-)$ for the conduction (valence) band and $J_a$ is the effective EAP scattering  strength.

In 1980, Pietronero \textit{et al.} \cite{Pietronero1980} used a tight-binding model to obtain this quasielastic rate and gave $J_a = \sqrt{3}a_0q_0J_0/2 \approx$ 9 - 12 eV with equal contributions from LA and TA modes  including only the gauge-field \cite{Park2014} or hopping energy \cite{Li2014} contribution. Since then, different EAP coupling models have been proposed to  extract $J_a$ from experimental data for graphene which ranges from 9 eV  to 29 eV \cite{Stauber2007,Vasko2007,Hwang2008,Bolotin2008,Efetov2010,Castro2010,Perebeinos2010,Kozikov2010,Mariani2010,Min2011,Kaasbjerg2012,Ochoa2012,Sohier2014}. A first-principles study \cite{Park2014}, which gave reasonable agreement with experimental data \cite{Efetov2010}, showed that $\tau_{TA}^{-1}(\epsilon_{sk}) \approx 2.5 \times \tau_{LA}^{-1}(\epsilon_{sk})$ in accord with Refs. \cite{Pietronero1980,Castro2010,Kaasbjerg2012,Ochoa2012} and with a previous first-principles analysis \cite{Borysenko2010}. This suggests that the gauge-field contribution is more important than the screened deformation potential, in agreement with Refs. \cite{Castro2010,Ochoa2012,Sohier2014}. A similar finding was reported in Ref. \cite{Li2014}, which gave $J_a^2=  E_1^2 + 9\beta^2\gamma_0^2\left( 1 + v_{LA}^2/v_{TA}^2 \right)/8 $. This is consistent with Refs.~\cite{Castro2010,Ochoa2012} with the assignment of $\vert{E_1}\vert = g_0/\epsilon (q)$  as the screened deformation potential [$\epsilon (q)$ is dielectric screening due to free carriers] and $\gamma_0 = 2\hbar v_F/\sqrt{3}a_0 \approx$ 3.1 eV. The terms  $E_1^2$,  $9\beta^2\gamma_0^2/8$, and $9\beta^2\gamma_0^2v_{LA}^2/8v_{TA}^2$ in $J_a^2$ are  contributions from the screened deformation potential due to LA phonons, the hopping energy terms (vector potentials) for LA modes, and that for TA modes, respectively. The  relative ratio of them is 1:4:10 \cite{Li2014}, which implies $\tau_{TA}^{-1}(\epsilon_{sk}) \approx 2 \tau_{LA}^{-1}(\epsilon_{sk})$, in agreement with  the first-principles study \cite{Park2014}. The single electron-phonon coupling parameter determined experimentally is not the screened (scalar) deformation potential $\vert{E_1}\vert$ but the effective deformation potential $J_a$ \cite{Li2014}. Recently, it has been shown \cite{Greenaway2019,Kumaravadivel2019} that $\rho_{TA}(\epsilon_{sk}) \approx 2 \rho_{LA}(\epsilon_{sk})$ implying $\tau_{TA}^{-1}(\epsilon_{sk}) \approx 2 \tau_{LA}^{-1}(\epsilon_{sk})$. Since $J_a$ depends on $v_{LA},\,v_{TA},\,\gamma_0\,(\mbox{or }v_F),\,\beta,\,\vert{E_1}\vert\,(\mbox{or } g_0/\epsilon(q))$, uncertainties in these parameters also contributed to the diverse values of $J_a$ mentioned above.

As we will discuss later, $J_a^2 ={ E_1^2 + 2B^2\left( 1 + v_{LA}^2/v_{TA}^2 \right) }$ \cite{Castro2010,Ochoa2012,Li2014} well explains the available data and reproduces other calculated results \cite{Sohier2014}, where $E_1$ is the screened deformation potential for LA phonons, $B$ is the electron-phonon coupling due to the hopping energy (or gauge field) terms, and $v_{LA} (v_{TA})$ is the sound velocity of LA (TA) phonons. %Eq.~(\ref{eq1}) gives  a reasonable agreement between predictions of first-principles studies and experimental data \cite{Castro2010,Ochoa2012,Park2014,Li2014}.

Currently, there exist a lot of controversies in the low $T$ behavior of EAP scattering rates. It has been believed that the low $T$  quasi-elastic scattering rate in graphene $\tau_{LT}^{-1}$ is proportional to $T^{n}$ with $n=2$ \cite{Viljas2010,Lucas2016}, $n=4$ \cite{Hwang2008,Mariani2008,Efetov2010,Mariani2010,Min2011,Sarma2011,Cooper2012}  or $n=6$ \cite{Hwang2008,Mariani2010}. The value  $n=4$  was claimed to be valid when $T <$ 10 K \cite{Morozov2008}, but it was not reproduced in Refs. \cite{Viljas2010,Sohier2014}. Inelastic EAP scattering rates have been evaluated numerically for graphene at finite temperature and carrier density via {\it ab initio} method \cite{Park2014,Sohier2014}. However, without analytical analysis, it is difficult to clarify the interplay of doping and temperature effects on EAP scattering processes and the range of validity of the commonly adopted quasielastic scattering rates at finite temperatures and doping densities.

Here we present a detailed analysis of inelastic acoustic phonon scattering rates, taking into account of both doping and temperature effects, which sheds light on the acoustic phonon scatterings in graphene, especially in the low $T$ regime that is still under debate \cite{Hwang2008,Mariani2008,Efetov2010,Viljas2010,Mariani2010,Min2011,Sarma2011,Cooper2012,Lucas2016}. Quasielastic \cite{Pietronero1980,Woods2000,Stauber2007,Vasko2007,Hwang2008,Bolotin2008,Chen2008,Efetov2010,Castro2010,Borysenko2010,Perebeinos2010,
Viljas2010,Kozikov2010,Mariani2010,Zou2010,Min2011,Sarma2011,Kaasbjerg2012,Cooper2012,Ochoa2012,Sohier2014,Li2014} and ansatz \cite{Efetov2010} equations of acoustic phonon scatterings are shown to be limiting cases of our inelastic equations, which well explain the experimental data \cite{Efetov2010,Dean2010,Kumaravadivel2019,Greenaway2019} and  agree with first-principles studies \cite{Park2014,Sohier2014} at different carrier densities for the whole range of $T$ considered. The nonlinearity in $T$ dependence of $\rho$ in both low-$T$ and high-$T$ regimes \cite{Tan2007,Morozov2008,Bolotin2008,Chen2008,DaSilva2010,Efetov2010,Castro2010,Ochoa2012,Park2014} are also discussed using inelastic equations and quasi-elastic limits extracted from them. Below we show that the main effect of inelastic EAP scattering is through the product of occupation number and electron distribution, even though the change of electron energy due to inelastic scattering is quite minor. The inelastic effect on EAP scattering rate becomes very significant at low temperature and high doping concentration when the chemical potential ($\mu$) is  much higher than $k_BT$. We also provide details of derivation for the semi-inelastic scattering rate, where we keep only the main effect of inelastic scattering on the product of phonon population and electron distribution. We then discuss the quasielatic limit and how the prediction deviates from the inelastic scattering results. The contributions from LA and TA modes at different temperatures and dopings are also analyzed. Finally, we  discuss the validity of Matthiessen's rule \cite{Matthiessen1864} and of the conventional determination of the effective deformation potential \cite{Li2010,Kaasbjerg2012,You2019}.

\section{Relations derived from momentum and energy conservation}

The 2D low-energy charged quasiparticles (i.e. electrons, electron holes) around a $K$ point in graphene can be described by a Dirac-like Hamiltonian $\mathcal{H} \Psi_{sk}(\bm{r}) = \epsilon_{sk} \Psi_{sk}(\bm{r})$ with $\mathcal{H} = \hbar v_F \bm{\sigma} \cdot \bm{k} = \hbar v_F ({\sigma}_x {k}_x + {\sigma}_y {k}_y)$, where  $\sigma$'s are the Pauli spin matrices, $v_F$ is the Fermi group velocity characterizing the $\pi$-band structure of graphene defined by $\hbar v_F = \sqrt{3}a_0\gamma_0/2$ with $a_0$ being the graphene's lattice constant and $\gamma_0$ being the hopping energy between the nearest neighbors, $\bm{\sigma} = (\hat{\sigma}_x, \hat{\sigma}_y)$ and $\bm{k}$ is the wave-vector. The electronic dispersion relation can then be obtained from the equation $\det(\mathcal{H} - \epsilon \mathcal{I}) = 0$, which gives $\epsilon_{sk} = s \hbar v_F k$, and the corresponding wave function is
$\Psi_{sk}(\bm{r}) = \chi_{sk} e^{i\bm{k}\bm{r}}/L$ with pseudospinor
$\chi_{sk} = \frac{1}{\sqrt{2}}
\left( \begin{array}{c} e^{-i\theta_k}	\\ s			\\
\end{array} \right)$ \cite{Ando2006}, where $L^2$ is the area of the graphene sheet, $s = \sgn(\epsilon_{sk})=+1 (-1)$ is the band index for $\pi^* (\pi)$ band, and $\theta_k = \atan(k_y/k_x)$ with $k^2 = k_x^2 + k_y^2$. Similar expressions can be worked out for the $K'$ point by flipping the sign of $\hat{\sigma}_y$. Because 2D Dirac systems are described by the same Hamiltonian, the forms of the eigenvalue $\epsilon_{sk} = s \hbar v_F k$ and  eigenfunction $\Psi_{sk}(\bm{r})$ still hold for any 2D Dirac materials with different values of $v_F$. Therefore, our full consideration of inelastic EAP scattering here is universal for all 2D Dirac materials.

The isotropic acoustic phonon dispersion relation is described by $\hbar\omega_a = \hbar v_a q$, where $a=LA (TA)$ labels the longitudinal (transverse) acoustic phonon. Because $v_a$ is much smaller than $v_F$ as in most of the known 2D materials, including graphene, for convenience, we use the dimensionless parameters $\zeta_a = (v_a/v_F)^2$ and $\gamma_a = 1/(1 - \zeta_a)$ with $0 < \zeta_a < 1$ and $\gamma_a > 1$. The momentum conservation law \cite{Ziman1960,Ashcroft1976,Lundstron2000} states that $\bm{q} = p (\bm{k}'-\bm{k})$, where $p = +1 (-1)$ corresponds to the absorption (emission) process, respectively, and gives
\begin{equation}
{k^\prime}^2 + k^2 - q^2 - 2 k^\prime k \cos\theta = 0,\label{eq2}
\end{equation}
where $\theta$ is the scattering angle between the initial momentum $\bm{k}$ and the final momentum $\bm{k}'$ and $\bm{q}$ is the transferred (i.e. absorbed or emitted) momentum. Using the  dispersion relation for $\epsilon_{sk}$ and $\hbar\omega_a$ above, we can rewrite Eq. (\ref{eq2}) as $\epsilon_{s'k'}^2+\epsilon_{sk}^2 - (\hbar\omega_a)^2/\zeta_a - 2 s's \epsilon_{s'k'}\epsilon_{sk}\cos\theta = 0$. Now we apply the energy conservation law \cite{Ziman1960,Ashcroft1976,Lundstron2000}: $\hbar\omega_a = p (\epsilon_{s'k'}-\epsilon_{sk})$ and obtain $\epsilon_{s'k'}^2 - 2\gamma_a\epsilon_{sk} \left( 1 - s's \zeta_a \cos\theta \right) \epsilon_{s'k'} + \epsilon_{sk}^2 = 0$. This quadratic equation of $\epsilon_{s'k'}$ (for a given $\epsilon_{sk}$) can be solved straightforwardly to give
\begin{eqnarray}
\epsilon_{sk}^p &=& \gamma_a \left[ \epsilon_{sk} \left( 1 - s' s \zeta_a \cos\theta \right) \right. \nonumber\\
&\pm & \left. \vert {\epsilon_{sk}} \vert  \sqrt{2\zeta_a \left( 1 - s' s \cos\theta \right) - \zeta_a^2 \left( 1-\cos^2\theta \right) } \right].\label{eq3}
\end{eqnarray}

It is  noted that $\epsilon^{\pm}_{sa}$ must have the same sign as  $\epsilon_{sk}$, otherwise the solution becomes unphysical. Namely, $s's=1$, which implies the inter-band  scattering is forbidden.
Thus, the $\pm$ in (\ref{eq3}) can be replaced by an index $p=\pm 1$, corresponding to the absorption ($+$) or emission ($-$) process. Thus, Eq.~(3) can be reduced to
\bea
\frac{\epsilon^p_{sa}}{\epsilon_{sk}} &=& 1+\gamma_a \left[ 2\zeta_a\sin^2\frac{\theta}{2} + p s \sqrt{4\zeta_a \sin^2\frac{\theta}{2}- \zeta_a^2\sin^2\theta} \right] \nonumber \\
&=& 1+\gamma_a c^p_a(y)  \label{eq4}
\eea
with $c^p_a(y)=2[\zeta_a y^2+p\sqrt{\zeta_a y^2-\zeta_a^2y^2(1-y^2)}]$ and $y\equiv \sin(\theta/2)$.

Eqs. (\ref{eq2}) - (\ref{eq4}) also give us the transferred momenta (i.e. the phonon momenta satisfying both the momentum and energy conservation laws) $q_{sa}^p$. For convenience, we define dimensionless auxialliary functions $ K^p_{sa}(\theta)=\frac{\epsilon^p_{sa}}{\epsilon_{sk}} $ and $Q_{sa}^p(\theta)=q_{sa}^p /k$. $Q_{sa}^p(\theta$) has two equivalent forms
\begin{subequations}
\label{eq5}
\begin{equation}
Q_{sa}^p(\theta) = \sqrt{ {K_{sa}^p(\theta)}^2 + 1 - 2 K_{sa}^p(\theta) \cos\theta },\label{eq5a}
\end{equation}
\begin{equation}
Q_{sa}^p(\theta) = ps (K_{sa}^p(\theta) - 1)v_F/v_a = ps \gamma_a c^p_{sa}(y) v_F/v_a.\label{eq5b}
\end{equation}
\end{subequations}
The transferred acoustic phonon energy is determined by $\hbar \omega_{sa}^p = \hbar v_a q_{sa}^p = \hbar v_a Q_{sa}^p(\theta) k = v_a s\epsilon_{sk}Q_{sa}^p(\theta)/v_F$. Similarly, by eliminating $\epsilon_{sk}^p$, we obtain inequalities $-1 \le \cos\theta_{\bm{kq}} = - pq/2\gamma_ak + sv_a/_F \le 1$, which give $\min(q_{sa}^p) = 0$ and $\max(q_{sa}^p) = 2(1 + psv_a/v_F)\gamma_ak$ with $\theta_{\bm{kq}}$ being the angle between $\bm{k}$ and $\bm{q}$.

The momentum and energy conservation laws lead to the constraints $\delta_{\bm{q},p(\bm{k'}-\bm{k})}$ and $\delta_{\epsilon} \equiv \delta(p (\epsilon_{s'k'} - \epsilon_{sk}) - \hbar v_a q) = \delta(p (\epsilon_{s'k'}-\epsilon_{sk}) - \frac{v_a}{v_F} \sqrt{\epsilon_{s'k'}^2+\epsilon_{sk}^2-2\epsilon_{s'k'}\epsilon_{sk}\cos\theta}) \equiv \delta{\left[ f_p(\epsilon_{s'k'}) \right]}$. Thus
\begin{eqnarray}
\delta_{\epsilon} &=& \frac{\delta(\epsilon_{s'k'}-\epsilon_{sk}^p)}{ \vert \frac{df_p(\epsilon_{s'k'})}{d\epsilon_{s'k'}} \vert_{\epsilon_{s'k'}=\epsilon_{sk}^p} }
%
%= \frac{\delta(\epsilon_{s'k'}-\epsilon_{sk}^p)}{ \vert p - \frac{v_a}{v_F} \frac{\epsilon_{s'k'} - \epsilon_{sk}\cos\theta}
%{ \sqrt{\epsilon_{s'k'}^2 + \epsilon_{sk}^2 - 2\epsilon_{s'k'}\epsilon_{sk}\cos\theta} } \vert_{\epsilon_{s'k'}=\epsilon_{sk}^p} }
%
= \frac{\delta(\epsilon_{s'k'}-\epsilon_{sk}^p)}{ d_{sa}^p(\theta) }\label{eq6}
\end{eqnarray}
with $d_{sa}^p(\theta) = \vert ps - \frac {v_a}{v_F} \frac{ K_{sa}^p(\theta) - \cos\theta }{ Q_{sa}^p(\theta) } \vert$ satisfying $d_{sa}^p(\theta) = d_{-sa}^{-p}(\theta)$. Note that Eq.~(\ref{eq6}) combines  two $\delta$-functions expressing momentum and energy conservations into a single $\delta$-function $\delta(\epsilon_{s'k'}-\epsilon_{sk}^p)/d_{sa}^p(\theta)$.

\begin{figure}[b]
\includegraphics[width=1.\linewidth]{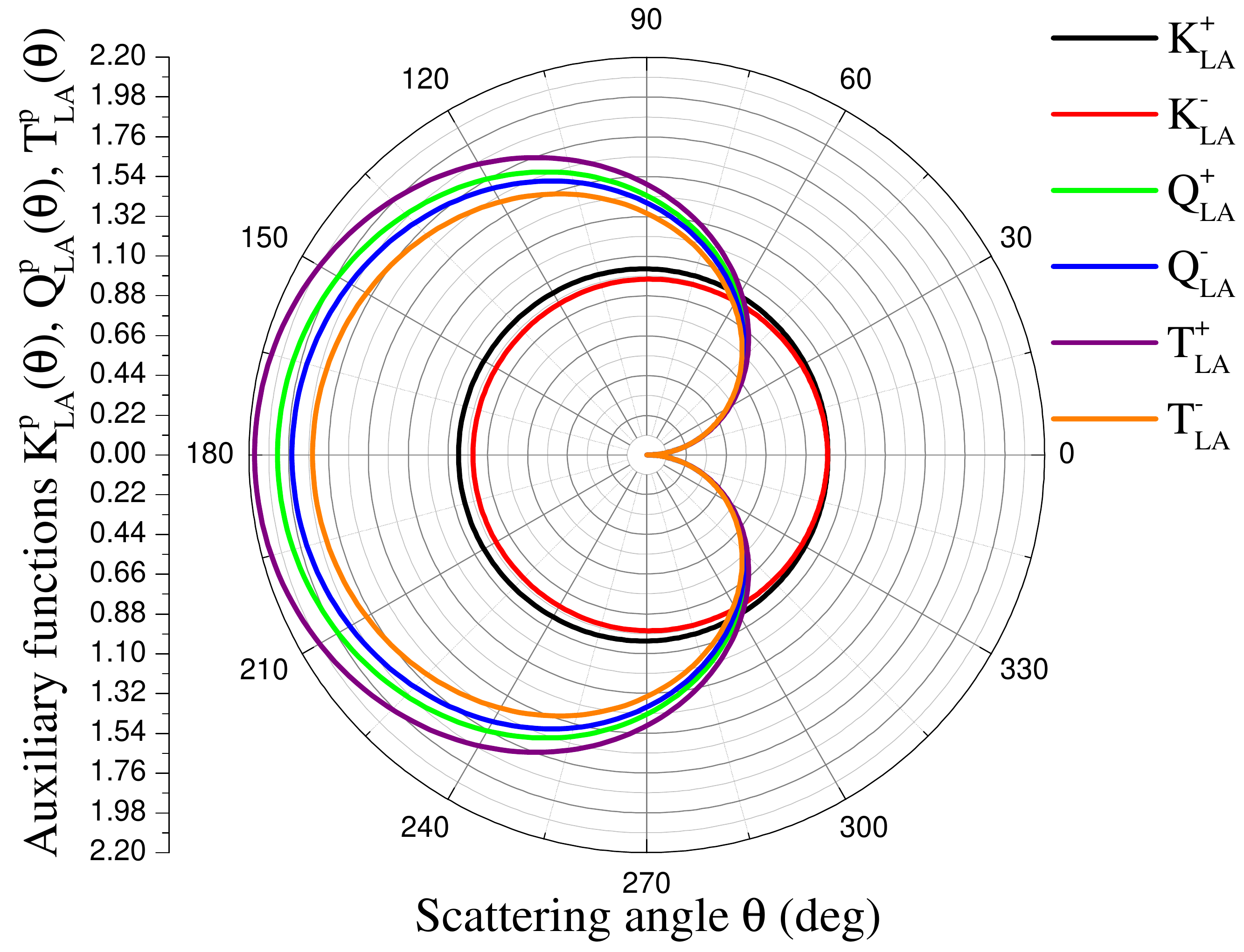}
\caption{\label{s1} The auxiliary functions $K_a^p(\theta)$, $Q_a^p(\theta)$ and $T_a^p(\theta)$ of graphene do not depend on $v_a$ and $v_F$ seperately but only on their ratio $v_a/v_F$ with $a=LA$.}
\end{figure}

%Auxiliary functions $K_{sa}^p(\theta)$ and $Q_{sa}^p(\theta)$ for $v_a/v_F=1/50$ (corresponding to LA mode) are shown in Fig.~\ref{s1}.
For simplicity, we only consider $s=+1$ and thus remove the $s$ index from the auxiliary functions in Fig.~\ref{s1}, which shows behaviors and symmetry properties of $K_{sa}^p(\theta) = \epsilon_{sk}^p/\epsilon_{sk}$, $Q_{sa}^p(\theta) = q_{sa}^p/k$ and $T_{sa}^p(\theta) = K_{sa}^p(\theta) Q_{sa}^p(\theta)/d_{sa}^p(\theta)$ for $a=LA$ of graphene. Obviously, $K_{sa}^p(\theta)$ are ellipses, not circles as their quasielastic counterparts. Moreover, they reflect the fact that,  $K_{+a}^+(\theta)$ for the absorption in the conduction band (which is the same as $K_{-a}^-(\theta)$ for the  emission in the valence band), increases from 1 up to $( 1 + \zeta_a ) \gamma_a + \sqrt{ ( 1 + \zeta_a )^2 \gamma_a^2 - 1 } \approx 1.04$, which is higher than the quasielastic counterpart of 1, for $\theta$ increasing from 0 to $\pi$ and decreases in a symmetrical way (i.e. from about 1.04 back to 1) for $\theta$ increasing from $\pi$ to $2\pi$. Inversely,  $K_{+a}^-(\theta)$ for the emission in the conduction band  decreases from 1 down to $( 1 + \zeta_a) \gamma_a - \sqrt{ ( 1 + \zeta_a)^2 \gamma_a^2 - 1 } \approx 0.96$, which is lower than the quasielastic counterpart of 1, for $\theta$ increasing from 0 to $\pi$ and increases symmetrically (i.e. from about 0.96 back to 1) for $\theta$ increasing from $\pi$ to $2\pi$. These facts show that the maximum variation in $K_{+a}^+(\theta)$ being equal to that in $K_{+a}^-(\theta)$ is about 4$\%$ and the maximum total variation between $K_{+a}^+(\theta)$ and $K_{+a}^-(\theta)$ is $2\sqrt{ ( 1 + \zeta_a )^2 \gamma_a^2 - 1 } \approx 8\%$. Concerning the absorption and emission processes in the conduction and valence bands, the same behaviors hold for $Q_{sa}^p(\theta)$; however, $q_{sa}^p$ are heart-shaped instead of elliptical \textit{orbits} as $\epsilon_{sk}^p$. Using $q_{sa}^p = Q_{sa}^p(\theta) k$ and $\max(q_{sa}^p) = 2(1 + ps v_a/v_F)\gamma_ak$, we observe that the maximum value of $Q_{+a}^-(\theta)$ is $2(1 - v_a/v_F)\gamma_a \approx 1.96$, which is lower than the quasielastic counterpart of 2, and of $Q_{+a}^+(\theta)$ is $2(1 + v_a/v_F)\gamma_a \approx 2.04$, which is higher than the quasielastic counterpart of 2. Similar to $K_{sa}^p(\theta)$, the maximum total variation between $Q_{+a}^+(\theta) = Q_{-a}^-(\theta)$ and $Q_{+a}^-(\theta) = Q_{-a}^+(\theta)$ is $4\gamma_av_a/v_F \approx 8\%$. In fact, it is true that $\sqrt{ ( 1 + \zeta_a )^2 \gamma_a^2 - 1 } = 2\gamma_a v_a/v_F$. Therefore, for heavy-doped 2D Dirac systems, graphene with high carrier densities for instance, besides the temperature effect, the doping effect must be taken into account properly.

\section{The static dielectric function used in the screened deformation potential due to doping}

Here we consider a graphene sheet encapsulated between an upper-layer material with a static dielectric constant $\epsilon_a$ and a lower-layer material with a static dielectric constant $\epsilon_b$ making an effective background static dielectric constant $\epsilon_r = (\epsilon_a+\epsilon_b)/2$ for the free-carrier screening in graphene \cite{Kotov2012,Kumaravadivel2019,Greenaway2019}. Because the transferred momenta in most cases are less than or equal to $2k$ [with $q_{sa}^p = Q_{sa}^p(\theta) k \approx 2k\sin(\theta/2)$] and the contribution from the screened deformation potential is much smaller than  the unscreened gauge field (or the hopping energy terms) \cite{Castro2010,Ochoa2012,Park2014,Li2014,Sohier2014,Kumaravadivel2019,Greenaway2019}, the static dielectric function in the random phase approximation (RPA) \cite{Kotov2012} can be evaluated at $k=k_F$ (implying $q \le 2k_F$) and we have
\begin{equation}
\epsilon (q) = \epsilon_r + \frac{g_sg_ve^2}{\hbar v_F} \frac{k_F}{q}.\label{eq7}
\end{equation}
Using the angular average of ${\bf q}$ we get $q = (2k_F/2\pi) \int d\theta \sin(\theta/2) = 4k_F/\pi$, which results in $\epsilon (q) = \epsilon_r + g_sg_v\pi e^2/4\hbar v_F = \epsilon_r + \pi e^2/\hbar v_F$ for $g_s=g_v=2$. Then the screened deformation potential becomes $|E_1| = g_0/\epsilon (q) \approx 2.54$ eV, which is in good agreement with Refs.~\cite{Castro2010,Ochoa2012,Li2014,Kumaravadivel2019}. Note that our approximation here gives $q = 4k_F/\pi$, which is greater than $q = k_F$ used in Ref. \cite{Castro2010} and smaller than $q = 2k_F$ in Ref. \cite{Sohier2014}. Moreover, $|E_1|$ becomes smaller when graphene is on or encapsulated between dielectric materials as a result of stronger screening \cite{Castro2010,Kumaravadivel2019}.

Interestingly, the energy and momentum conservation laws result in the selection rule $ss_p = 1$, which  in turn leads to suppressed forward- and backward-scattering rates via the chiral term $\chi_{\theta} = \vert \chi_{sk}^{\dagger} \chi_{s'k'} \vert^2 = \vert e^{i(\theta_k - \theta_{k'})} + ss' \vert^2/4 = (1 + \cos\theta)/2$ in the screened deformation potential contribution.

\begin{figure}[b]
\includegraphics[width=1.\linewidth]{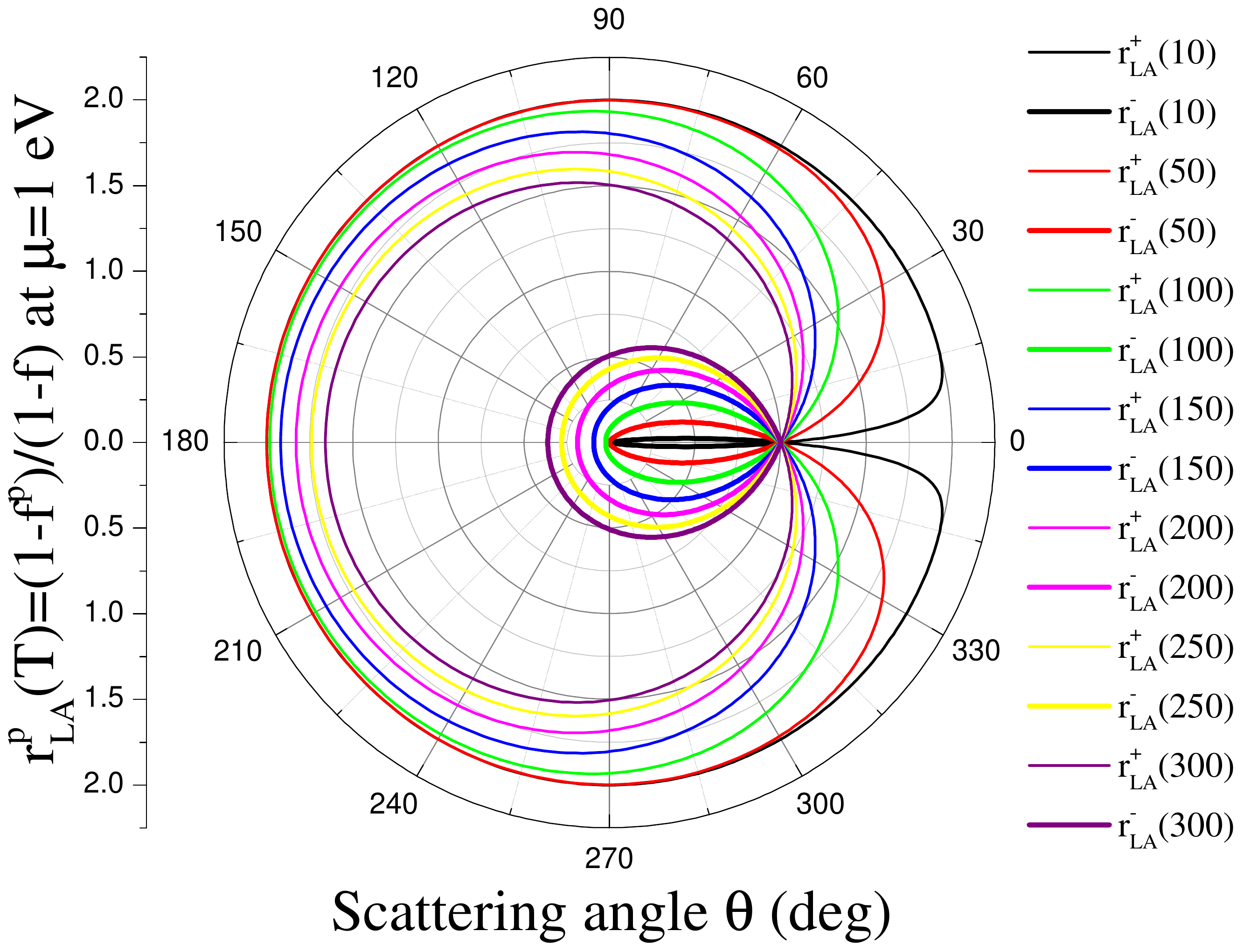}
\caption{\label{s2} The ratio between $1-f(\epsilon_{sk}^p)$ and $1-f(\epsilon_{sk})$ as a function of $\theta$ and temperature varying from 10 to 300K at $\mu$=1 eV. The thinner and thicker curves show the absorption and emission processes, respectively, with $a=LA$.}
\end{figure}

\section{The energy-dependent inelastic EAP scattering rates}

The full momentum relaxation rate due to inelastic scattering  by acoustic phonons  is given by \cite{Li2014}
\begin{eqnarray}
 \frac 1 {\tau^s_{\bf_k}} &=& \frac {1} {2\rho_m} \int \frac {k'dk'}{2\pi}\sum_{a,p}\int d\theta(1-\cos \theta)\frac{ 1 - f(\epsilon_{s'k'})} { 1 - f(\epsilon_{sk}) }\nonumber \\
 &\times& q F^a_{\bf k',\bf k} [ N^{(a)}_{\bf q} \frac{1}{2} - \frac{p}{2}  ] \delta(\epsilon_{sk}-\epsilon_{s'k'}+p\hbar\omega^{a}_{q}), \label{rate}
\end{eqnarray}
where $s=\pm 1$ labels whether the electron is in upper or lower Dirac cone before scattering. $p=\pm 1$ indicates the  absorption ($+$) and emission ($-$) processes, $q=|{\bf k'}-{\bf k}|$,  $\theta=\theta_{\bf k'}-\theta_{\bf k}$, $\rho_m$ is the mass density, and  $N^{(a)}_{\bf q}$ denotes the phonon occupation number. $F^{(LA)}_{\bf k',\bf k}= \frac {1 }  {v_{LA}}|E_1\cos(\theta/2)+B \cos (3\theta/2+3\theta_{\bf k})|^2$ and $F^{(TA)}_{\bf k',\bf k}= \frac {1 }  {v_{TA}}|B \sin (3\theta/2+3\theta_{\bf k})|^2$. $E_1$ is the screened deformation potential for LA phonons and $B= {3\beta \gamma_0}/4$ \cite{Li2014}. If we take an average  over  $\theta_{\bf k}$ for the $\theta_{\bf k}$-dependent terms in (\ref{rate}), we get the same equation for the angle-average rate with $F^{(LA)}_{\bf k',\bf k}$ replaced by $ F_a(\theta)$, where $F_{LA}(\theta)= \frac {1} {v_{LA}}[E^2_1\cos^2(\theta/2)+\frac 1 2 B^2]$ and $ F_{TA}(\theta)=\frac {1}  {2v_{TA}} B^2$.

By averaging over the orientation of {\bf k} in (\ref{rate}), we obtain the energy-dependent relaxation rate at a finite $T$ for any chemical potential ($\mu$) as
\begin{eqnarray}
&\tau^{-1}_{in}(\epsilon_{k})& = \frac{ \Upsilon (\epsilon_{k}) }{ 1 - f(\epsilon_{k}) } \int {d\theta} ( 1-\cos \theta) \sum_{a,p}  \nonumber\\
&\times & D^p_{a}(\theta) \left(  N^{(a)}_{\bf q} + \frac{1}{2} - \frac{p}{2} \right) \left[ 1 - f(\epsilon^p_{k}) \right],\label{rate0}
\end{eqnarray}
where $\Upsilon(\epsilon_{k})= {\epsilon_{k}^2 }/ {4\pi \hbar^3 v_F^3 \rho_m}$, $D^p_{a}(\theta) = F_a(\theta) T^p_{a}(\theta)$,
$T^p_{a}(\theta)=K^p_{a}(\theta) Q^p_{a}(\theta)/d^p_{a}(\theta)$, and $d^p_{a}(\theta) =[c^p_a(y)-4\zeta_ay^2]/ |\gamma_a c^p_a(y)| $. To the first order of $v_a/v_F$, we have $K^p_{a}(y)=1+2p y v_a/v_F$, $Q^p_{a}(y)=2y$, and  $d^p_{a}(y) \approx (1- 2pyv_a/v_F)$. $N^{(a)}_{\bf q}$ and $f(\epsilon_k)$ are the Bose-Einstein and Fermi-Dirac distribution functions, respectively.

With $\tau^{-1}(\epsilon_{k})$ given in Eq.~(\ref{rate0}), the conductivity $\sigma$ can be calculated according to \cite{Ashcroft1976}
\be
\sigma = {e^2} \int \frac{k dk}{\pi} v_F^2 \tau(\epsilon_{k})[-d f(\epsilon_{k})/d\epsilon_{k} ],\label{tau}
\ee
where $-d f(\epsilon_k)/d\epsilon_k$  can be approximated by $\delta(\epsilon_k-\mu)$ when $\tau(\epsilon_{k}) |\epsilon_{k}| $ is slow varying over the range of $k_BT$.

Because $-{d f(\epsilon_{k})}/{d\epsilon_{k}}\vert_{T>0} = f(\epsilon_{k}) \left[ 1 - f(\epsilon_{k}) \right]/k_BT \approx  \delta(\epsilon_{k} - \mu)$, the scattering rate
${1}/{\tau(\epsilon_{k})}$ is often replaced by ${1}/{\tau(\mu)}$ in practical applications. As $\epsilon_{k} \rightarrow \mu$, we have
$1-f(\epsilon^p_{k}) = e^{px^p_{a}}/(e^{px^p_{a}}+1)$, $1-f(\epsilon_{k}) \rightarrow 1-f(\mu) = 1/2$ and
$ N^{(a)}_{\bf q} + 1/2 - p/2 = p/(e^{px^p_{a}}-1)$, where $x^p_{a}\equiv \hbar\Omega^p_{a}/k_BT$ with $\hbar\Omega^p_a \equiv |\epsilon_{k} (K^p_a(\theta)-1)| = |\epsilon_{k} \gamma_a c^p_a(y)|=v_a \vert\mu\vert Q_{a}^p(\theta)/v_F $. Thus we have
\be
 \left( N^{(a)}_{\bf q} + \frac{1}{2} - \frac{p}{2} \right) \frac{1 - f(\epsilon^p_{k}(\theta))}{1 - f(\epsilon_{k})} = \csch(x^p_{a})
\ee
for both $p=\pm 1$, where  $\csch(x)$ denotes the hyperbolic cosecant function. Finally, for finite $\mu$, we obtain
\be
\frac 1 {\tau_{in}(\mu)} = \Upsilon(\mu) \int  d\theta ( 1-\cos \theta)  \sum_{a,p}  D^p_{a}(\theta) \csch(x^p_{a}).\label{rate1}
\ee
Eqs.~(\ref{rate0}) and (\ref{rate1}) are our main results which show how doping and temperature effects come into play in the EAP scattering rates in 2D Dirac materials.
If $k_BT \gg \hbar \Omega^p_a$, we use $\csch(x) \approx 1/x-x/6$ to obtain
\begin{equation}
\frac 1 {\tau_{HT}(\mu)} = \Upsilon(\mu)  \int {d\theta}(1-\cos \theta) \sum_{a,p} D^p_{a}(\theta) (\frac 1 {x^p_{a}}-\frac {x^p_{a}} 6).\label{rHT}
\end{equation}

Figure~\ref{s2} shows the ratio $r_a^p(T) = \frac{1-f(\epsilon_{k}^p)}{1-f(\epsilon_{k})}$ as a function of $\theta$ and temperature varying from 10 to 300K at $\mu$=1 eV for $a=LA$ of graphene. Note that $r_a^+(T) > 1$ and $r_a^-(T) < 1$, which is very different from the quasielastic approximation, $r_a^p(T) = 1$.

\section{The energy-dependent semi-inelastic EAP scattering rates}

For graphene, we have $v_a/v_F \ll 1$. We can take the limit [$k' \rightarrow k$ and $q  \rightarrow 2 k \sin(\theta/2)$] and $x^p_{a}\approx (\Theta^a_F/T)\sin(\theta/2)$ with $\Theta^a_F\equiv 2 \hbar v_a k_F /k_B$ being a characteristic temperature.  Thus, we can  get a "semi-inelastic" equation by simply replacing $D^p_a(\theta)$ in Eq.~(\ref{rate0}) with $2\sin(\theta/2)F_a(\theta)$ and sum over $p$ to get a factor of 2. Our semi-inelastic equation can reproduce all results of inelastic scattering given  in Eq.~(\ref{rate0}) with $\sim$ 1\% error.
In the high-T regime, we can deduce from Eq.~(\ref{rHT}) a quasi-elastic limit, which contains an extra term in comparison with the quasielastic results derived in previous studies \cite{Sarma2011,Li2014}. We have
\begin{equation}
\tau^{-1}_{HT}(\mu) \approx {\Upsilon(\mu)} \frac {k_B T}{\mu} [D_t-\frac {G_t \mu^2}{3(k_BT)^2}] ,\label{rHT1}
\end{equation}
where
\bea
D_t &=& 2 v_F \int  {d\theta} (1-\cos \theta)  \sum_a  F_a(\theta)/ v_a \nonumber  \\
&=& \pi v_F[(E_1^2+ 2 B^2)/v_{LA}^2  +  2B^2/ v_{TA}^2 ]  \label{eqDa}
\eea
\bea
G_t  &=& \frac 2 {v_F} \int  {d\theta}(1- \cos \theta)^2\sum_a  F_a(\theta) v_a \nonumber  \\
&=&  \frac  {\pi} {v_F}(E_1^2 + 6B^2)   \label{eqGa}
\eea
The leading term in (\ref{rHT1}) is the same as (\ref{eq1}) for $\epsilon_k = \mu$, which was derived in Refs.~\cite{Castro2010,Ochoa2012,Li2014} with explicitly $J_a^2=v_{LA}^2 D_t/\pi v_F$. The second term in (\ref{rHT1}) provides a correction to (\ref{rHT1}) which is significant when $\mu$ is comparable to $k_B T$.

In general, since $\zeta_a \ll 1$ for $a=LA,\,TA$  in graphene, we can take the limits $\zeta_a \rightarrow 0$ and $\gamma_a \rightarrow 1$ in the auxiliary functions $K_{a}^p(\theta),\,Q_{a}^p(\theta),\,d_{a}^p(\theta)$, and $T_{a}^p(\theta)$ and we get $K_{a}^p(\theta) \rightarrow 1$, $Q_{a}^p(\theta) \rightarrow 2\sin(\theta/2)$, $d_{a}^p(\theta) \rightarrow 1$, and $T_{a}^p(\theta) \rightarrow Q_{a}^p(\theta) \rightarrow 2\sin(\theta/2)$. As a result, the inelastic EAP scattering rates given by Eq.~(\ref{rate0}) reduces to the semi-inelastic EAP scattering rates at $\epsilon_{k}$. We have
\bea
\frac{1}{\tau_{si}(\epsilon_{k})} &=& 4 \Upsilon (\epsilon_{k}) \int d\theta \sin^3  \frac \theta 2 \sum_{a,p} {F_a(\theta)} \nonumber  \\
&\times& \left( N_{a}(\theta) + \frac{1}{2} - \frac{p}{2} \right) \frac{1 - f(\epsilon_{k}^p)}{1 - f(\epsilon_{k})}.\label{eq10}
\eea

\begin{figure}[b]
\includegraphics[width=1.\linewidth]{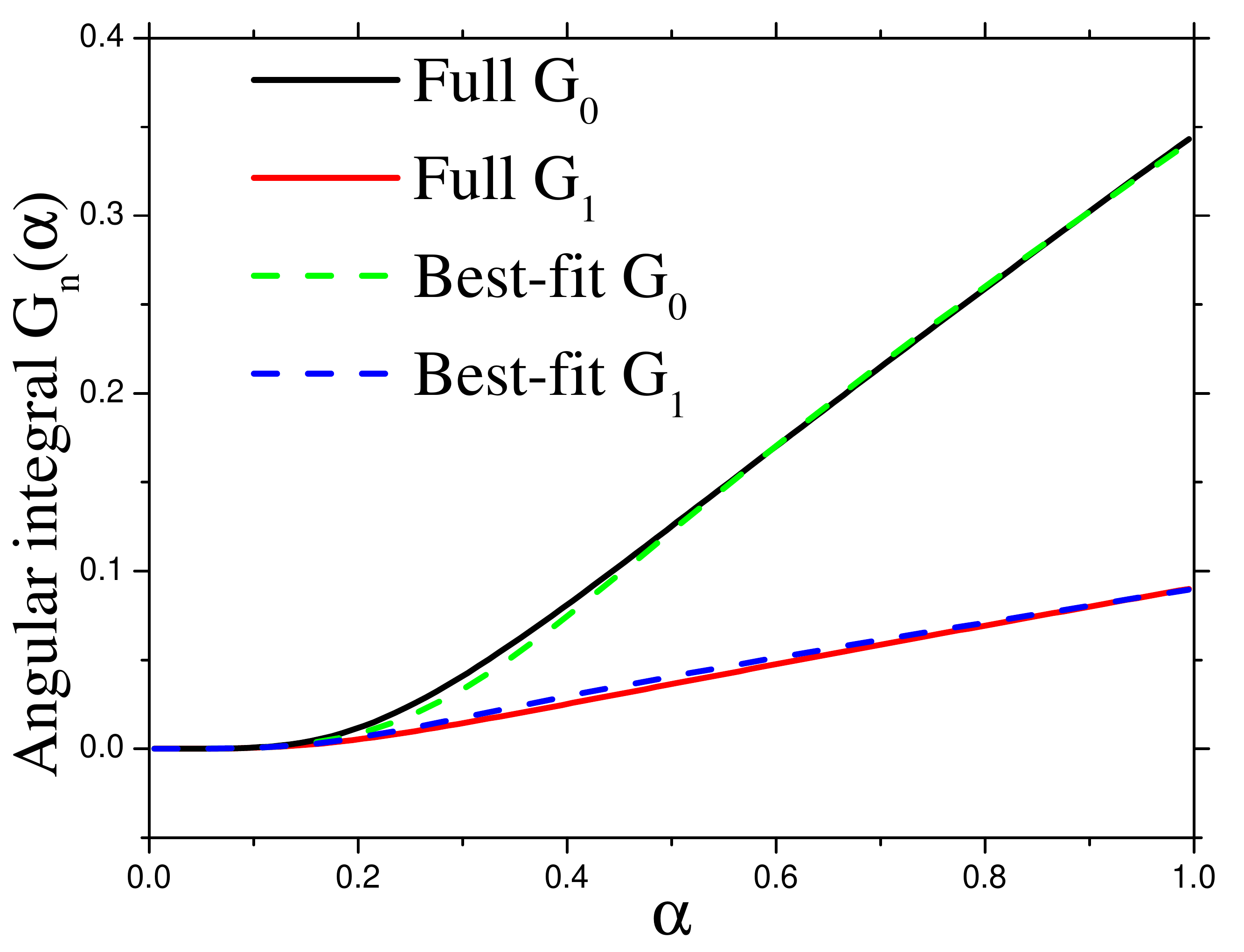}
\caption{\label{s3} Comparison of $G_n(\alpha)$ calculated by carrying out the integral in Eq.~(\ref{Gn}) and that calculated by using the analytic expression  $6\alpha_a^4/(1+c_n\alpha_a^3)$ with $c_0=16.5$, and $c_1=65.7$.}
\end{figure}

The phonon occupation number is now given by $N_a(\theta) = 1/\left( e^{\hbar\omega_a/k_BT} - 1 \right)$ with  $\hbar\omega_a = 2 v_a \vert\epsilon_{k}\vert\sin(\theta/2)/v_F$.
Eq.~(\ref{mu0}) can be evaluated numerically. We found that ${\cal R}$ is almost 1 (with $\sim 1$\% error) for any value of $k_BT$ and $\epsilon_{k}$ as implied in Figs.~\ref{fig2} and \ref{s7}. Thus, the quasielastic approximation given by Eq.~(\ref{eq1}) works extremely well for $\mu=0$ and remains a good approximation as long as
$\hbar\omega_{a}^p/k_BT \le {2v_a \vert\epsilon_k\vert}/{v_F k_BT} \ll 1$ or $\vert\mu\vert \ll {v_F k_BT}/{2v_a} \approx 25 k_BT$ for graphene.
At $\epsilon_{k}=\mu$, we define $\alpha_a=T/\Theta^a_F$; then we have
\bea
&& \frac{1}{\tau_{si}(\mu)} = 8 \Upsilon (\mu)\int d\theta \sin^3  \frac \theta 2  \sum_a {F_a(\theta)} \csch \frac {\hbar\omega_a}{k_BT}
\nonumber  \\
%&&= 16 \Upsilon(\mu) \int_0^1 dy \sum_a\frac {y^3\csch(y/\alpha_a)}{v_a \sqrt{1-y^2}} [2\delta_{a,LA}E_1^2 (1-y^2)+ {B^2}] \nonumber  \\
&& = 32 \Upsilon(\mu) \sum_{a} \frac{1}{v_a} [G_1(\alpha_a)2\delta_{a,LA}E_1^2 + G_{0}(\alpha_a){B^2}],\label{eq11}
\eea
where
\be G_n(\alpha_a)=\int_0^1 dy (1-y^2)^{n-1/2} y^3/(e^{y/\alpha_a}-e^{-y/\alpha_a}).\label{Gn}
\ee
$G_n(\alpha_a)$ (for $n=0,1$) can be well fitted by an analytic expression of the form $6\alpha_a^4/(1+c_n\alpha_a^3)$ and we get $c_0=6/G_0(1)-1=16.5$, and $c_1=6/G_1(1)-1=65.7$  as shown in Fig.~\ref{s3}.

Thus, we obtain a simple analytic expression for the semi-inelastic scattering rate
\bea \label{rsi}
\frac 1 {\tau_{si}(\mu)}& = &  {12 \Upsilon(\mu)} \left\{ \frac{(v_F k_B T)^4}{\mu^4 v_{LA}^5}
\left[ \frac{2E_1^2}{1+c_1 \alpha_{LA}^3} +\frac {B^2} {1+ c_0\alpha_{LA}^3} \right] \right.\nonumber  \\
&+ & \left. \frac{(v_F k_B T)^4}{\mu^4 v_{TA}^5}\frac {B^2} {1+ c_0\alpha_{TA}^3} \right\}
\eea

Our semi-inelastic expression contains a leading term of $T^4$  behavior at low $T$ and it approaches linear $T$ behavior at high $T$. This expression gives nearly the same results as the full inelastic expression given in Eq.~(\ref{rate1}) for graphene at practically all $T$ and $\mu$ of interest.
In the low-T limit ($\alpha_a \rightarrow 0$),  the above equation reduces to
\be
\frac 1 {\tau_{LT}(\mu)} = {12 \Upsilon(\mu)} \frac{(v_F k_B T)^4}{\mu^4 v_{LA}^5} \left[ 2E_1^2+B^2 \left( 1+\frac{v_{LA}^5}{v_{TA}^5} \right) \right]. \label{rLT}
\ee

The semi-inelastic EAP rates given by Eqs.~(\ref{eq10}) and (\ref{eq11}) can reproduce the inelastic EPS rates given by Eqs.~(\ref{rate0}) and (\ref{rate1}) with a high precision ($\sim$ 99\%). Moreover, the semi-inelastic equations are simpler than the full inelastic ones and much better than the quasielastic approximation given below.

\begin{figure}[b]
\includegraphics[width=1.\linewidth]{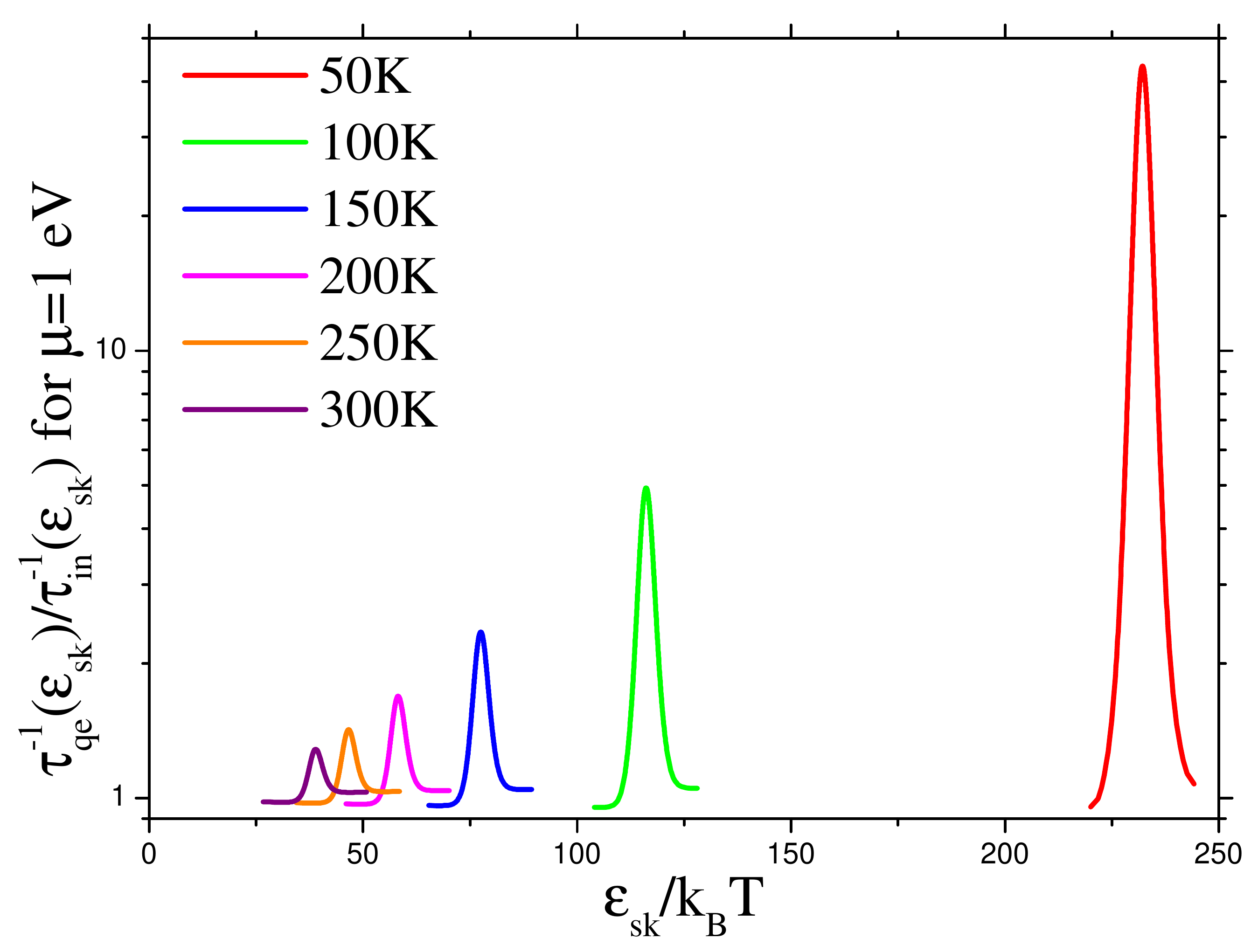}
\caption{\label{s4} The ratio between the quasielastic $\tau_{qe}^{-1}(\epsilon_{sk})$ and inelastic $\tau_{in}^{-1}(\epsilon_{sk})$ scattering rates as a function of $\epsilon_{sk}/k_BT$ and temperature varying from 50 to 300K at $\mu$=1 eV. }
\end{figure}

\section{The energy-dependent quasielastic EAP scattering rates}

Here we describe the energy-dependent quasielastic EAP scattering rates  at  finite $T$. This is derived by setting
$f(\epsilon_{k}+\hbar\omega_a) \approx f(\epsilon_{k}-\hbar\omega_a)\approx f(\epsilon_{k})$, $\hbar\omega_a = 2(v_a/v_F) \vert\epsilon_{k}\vert\sin(\theta/2)$, and $2N_a(\theta) + 1 = \frac{\exp(\hbar\omega_a/k_BT) + 1}{\exp(\hbar\omega_a/k_BT) - 1}$. We get
\be
\frac{1}{\tau_{qe}(\epsilon_{k})}= 4\Upsilon (\epsilon_{k}) \int d\theta \sin^3 \frac \theta 2
 \sum_a {F_a(\theta)} \frac{e^{\hbar\omega_a/k_BT} + 1}{e^{\hbar\omega_a/k_BT} - 1},\label{eq12}
\ee
which reproduces Eq.~(\ref{eq1}) $\tau_{HT}^{-1}(\epsilon_{k}) = J_a^2\vert\epsilon_{k}\vert k_BT/4\rho_mv_{LA}^2\hbar^3v_F^2$ for $\hbar\omega_a/k_BT \ll 1$. Fig.~\ref{s4} shows the ratio between the quasielastic $\tau_{qe}^{-1}(\epsilon_{sk})$ and inelastic $\tau_{in}^{-1}(\epsilon_{sk})$ as a function of $\epsilon_{sk}/k_BT$ for temperature varying from 50 to 300K with $\mu$=1 eV. As seen in the figure, the ratio peaks at $\epsilon_{sk}=\mu$ with a maximum deviate substantially from 1 except at high temperatures ($T$=300K or higher).  For $\epsilon_{sk}$ far away from $\mu$, the  ratio quickly approaches 1. However, since $\tau_{in}^{-1}(\mu)$ gives the dominant contribution for transport, the inelastic equation is needed to calculate the transport  properties accurately at low temperatures and high dopings. Only at high temperatures or low dopings, the quasielastic approximation is valid; that is when $\vert\mu\vert \ll {v_F k_BT}/{2v_a} \approx 25 k_BT$ for graphene.

For the intrinsic case ($\mu=0$), it can be shown that the quasielastic  limit of (\ref{rate0}) reduces to Eq.~(\ref{eq1}). In deed, the ratio of the inelastic scattering rate of  Eq.~(\ref{rate0}) to the "high-T" quasielastic limit, $\tau_{HT}^{-1}(\epsilon_k)$ of  Eq.~(\ref{eq1}) is
 \bea  \label{mu0}
&& {\cal R} = \frac{\tau_{HT}(\epsilon_k)}{\tau_{in}(\epsilon_k)} = \frac{\vert\epsilon_k\vert}{D_t k_BT} \int {d\theta} {(1-\cos \theta)}  \nonumber \\
&& \times  \sum_{a,p} D^p_{a}(\theta) \left( N_a(\theta) + \frac{1}{2} - \frac{p}{2} \right)  \frac{ 1 - f(\epsilon^p_{k}(\theta))}{1 - f(\epsilon_{k})}.
\eea

\section{EAP scattering rates in graphene}

\begin{figure}[t!]
\includegraphics[width=1.\linewidth]{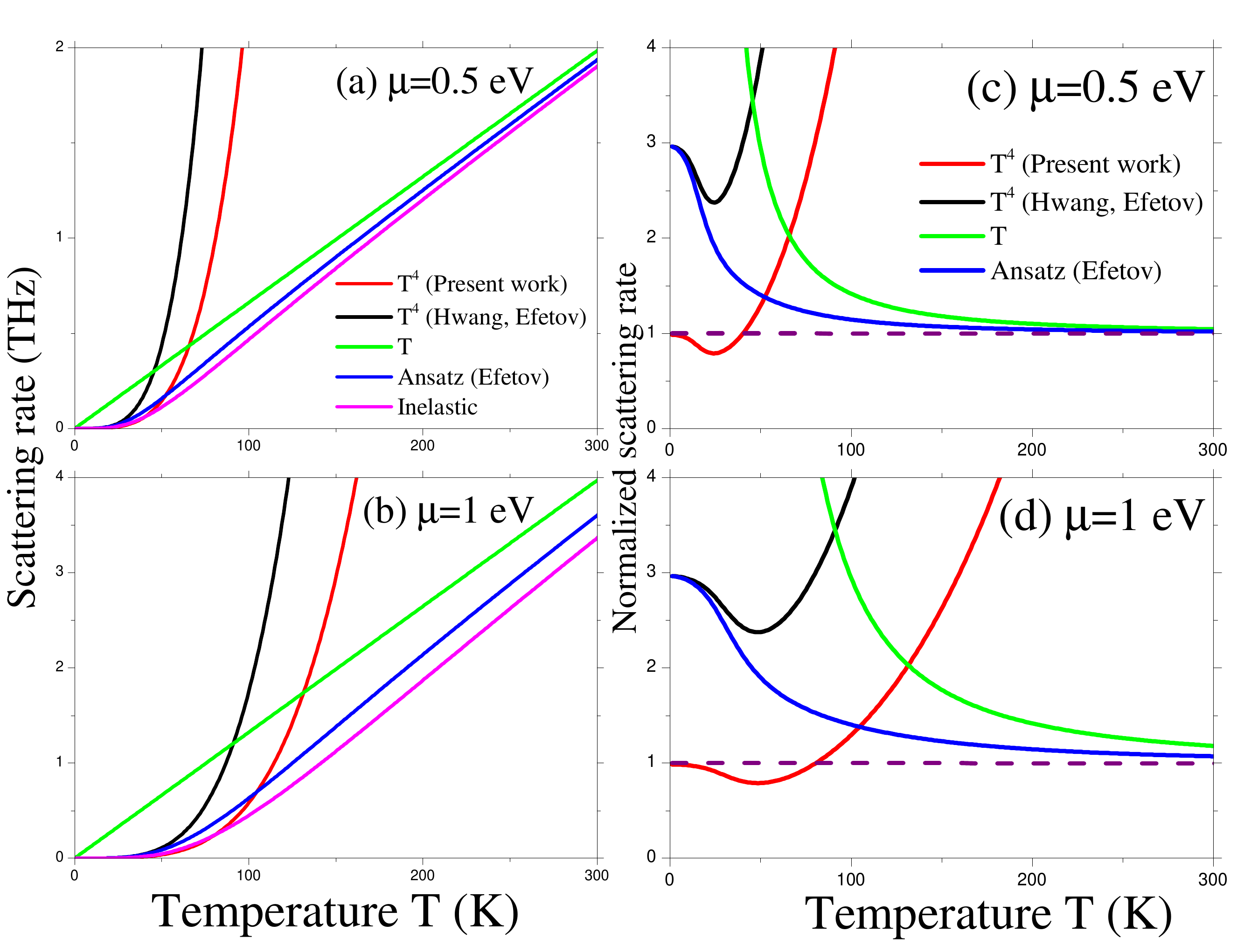}
\caption{\label{fig1}
Calculated  EAP scattering rates of graphene for (a) $\mu$ = 0.5 eV and (b) $\mu$ = 1 eV. (c) and (d)  the same rates from (a) and (b), normalized to $1/\tau_{in}(\mu)$ as given by Eq. (\ref{rate1}). The $T^4$ rate \cite{Hwang2008,Efetov2010} is $1/\left<\tau_{LT}(\mu)\right> \approx 4!\zeta(4)J_a^2 (k_BT)^4/2\pi\rho_mv_{LA} \vert\mu\vert k_F (\hbar v_{LA})^4$, the ansatz rate \cite{Efetov2010} is $1/\left<\tau(\mu)\right> = 8J_a^2\mu^2 f_s(z)/\pi\rho_mv_{LA}\hbar^3v_F^3$ with $f_s(z) = \int_0^1 z u^4 \sqrt{1-u^2} \exp(zu) du/[\exp(zu)-1]^2$, $z = \Theta_F^{LA}/T$, and the linear-in-$T$ rate is calculated by Eq. (\ref{eq1}) at $\mu$. The dash curves in panels (c) and (d) are the ratios between $1/\tau_{si}(\mu)$ given by Eq. (\ref{eq11}) and $1/\tau_{in}(\mu)$. Here we use $\beta=3$.}
\end{figure}

We now compare results predicted by the current model with those derived previously \cite{Hwang2008,Efetov2010}. Because the scattering rates vary in orders of magnitude, it is  illustrative to also compare the ratio of results predicted by various approximations to that of the full inelastic-scattering result given by Eq.~(\ref{rate1}) (corresponding to Eq.~(\ref{rate0}) at $\epsilon_k=\mu$). The results are  shown in Fig.~\ref{fig1}. Throughout the paper, we use $v_F = 1.0 \times 10^6$ (m/s), $v_{LA} = 2.0 \times 10^4$ (m/s), $v_{TA} = 1.3 \times 10^4$ (m/s), $\rho_m = 7.6 \times 10^{-7}$ (Kg/m$^2$) \cite{Kumaravadivel2019,Greenaway2019}, $g_0$ = 20 (eV) \cite{Suzuura2002,Castro2010,Kogan2014}, and $\beta = 2.75 \sim 3$ \cite{Castro2010,Ochoa2012,Li2014} depending on samples. It is seen that the high-$T$ scattering rate, $\tau_{HT}^{-1}$ predicted by Eq.~(\ref{eq1}) (green curve) merge into the dashed line (full inelastic result) slowly (from above). Note that, when we remove the screened deformation potential  (i.e. let $E_1\rightarrow 0$ ), $J_a \rightarrow \sqrt{2}B\sqrt{1+v_{LA}^2/v_{TA}^2}$ which implies the acoustic gauge field in Ref.~\cite{Sohier2014} $\beta_A = B/\sqrt{2} = 3\beta\gamma_0/4\sqrt{2} \approx 4.5 \sim 4.9$ when $\beta = 2.75 \sim 3$ is used. These values are in excellent agreement with the GW and fitted values given in Ref.~\cite{Sohier2014}. The effect of removing $E_1$ is discussed in Fig.~\ref{s6} below.

The low-$T$ scattering rate, $\tau_{LT}^{-1}$ given by Eq.~(\ref{rLT}) (red curve) is indistinguishable from the inelastic result for $T<100 K$, while the result based on the $T^4$ formula, $\tau_{LT}^{-1}=4!\zeta(4)J_a^2 (k_BT)^4/2\pi\rho_m v_{LA} \mu k_F (\hbar v_{LA})^4$ as given in \cite{Hwang2008,Efetov2010} (black curve) deviate from the full inelastic result by a factor $\sim$ 3 at $T$=1K and much more than 3 for $T$ above 50 (100) K for $\mu=$ 0.5 (1) eV. It is interesting to note that the ansatz formula (blue curve), $\tau^{-1}(\mu) = 8J_a^2\mu^2 f_s(z)/\pi\rho_m v_{LA}\hbar^3v_F^3$ \cite{Efetov2010} with $f_s(z) = \int_0^1 z u^4 \sqrt{1-u^2} \exp(zu) du/[\exp(zu)-1]^2$ and $z = \Theta_F^a/T$ as used in \cite{Efetov2010} can match  the full inelastic result well for $T> 200 K$, but deviate significantly (also by a factor $\sim$ 3) as $T$ approaches 0. This factor of $\sim$ 3 difference is caused by the approximation used in previous works in which the factor $[1 - f(\epsilon^p_k)]/[1 - f(\epsilon_k)]$ was replaced by 1 that turns out to be problematic at low-$T$. Finally, results from our semi-inelastic formula given in Eq.~(\ref{eq11}) (dash curves in (c) and (d)) match the full inelastic results nicely (with error $\sim$ 1\%) in the entire range of T and $\mu$.

\section{Resistivity due to EAP scattering in graphene}

\begin{figure}[t!]
\includegraphics[width=1.\linewidth]{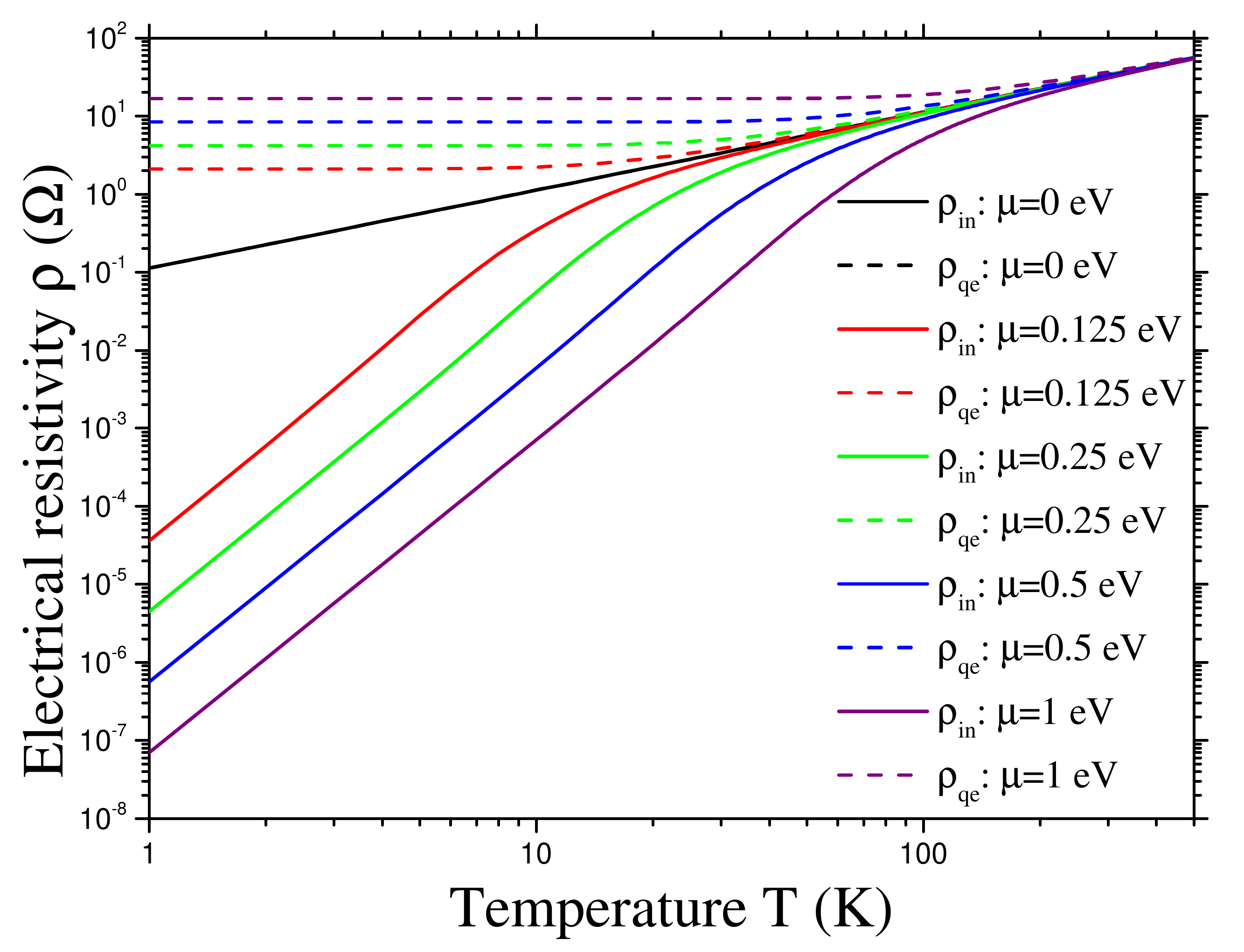}
\caption{\label{fig2}
Graphene's electrical resistivity is calculated from the inelastic (the solid curves) and quasielastic (the dash curves) scattering rate at $\mu$=0 eV (the black curves), $\mu$=0.125 eV (the red curves), $\mu$=0.25 eV (the green curves), $\mu$=0.5 eV (the blue curves), and $\mu$=1 eV (the purple curves). Here we use $\beta=3$. The linear plot is displayed in Fig.~\ref{s7}.}
\end{figure}

Using Eq.~(\ref{tau}) (without replacing $-df(\epsilon)/d\epsilon$ by a delta function) we can evaluate the resistivity $\rho = 1/\sigma$ and compare results of our full inelastic model with those obtained by quasielastic model in a log-log plot in Fig.~\ref{fig2}. Here, we have performed the integral over $k$ in Eq.~(\ref{tau}) numerically but keeping the full energy dependence of $\tau(\epsilon_k)$. Had we approximated $ -df(\epsilon)/d\epsilon$   by $\delta(\epsilon-\mu)$ in $\tau(\epsilon_k)$ as in common practice, the calculated resistivity would have been about 30\% lower at low temperatures as shown in Fig.~\ref{s5}. We see that for intrinsic case ($\mu=0$), the quasielastic model works extremely well as anticipated (since $\rho \propto 1/\tau$ in this case). However, at finite $\mu$, significant deviation (up to 6 orders of magnitudes) occurs. However, if we use the semi-inelastic expression in Eq.~(\ref{eq11}), the predicted resistivities still match the full inelastic results nearly perfectly (with unobservable difference not displayed in this plot). Interestingly, at $T<100K$, the full inelastic calculations predict that the resistivity decreases as doping increases, which is opposite to the results predicted by the quasielastic model and the common perception. Such a prediction, however, is consistent with experimental findings \cite{Tan2007a}. Here, we only considered the contribution from EAP scattering alone. Thus, at high dopings we find that the resistivity can be extremely low ($<10^{-4} \Omega$) at low temperatures. However, in realistic samples other mechanisms such as defect and carrier-carrier scatterings must be considered.

\begin{figure}[b]
\includegraphics[width=1.\linewidth]{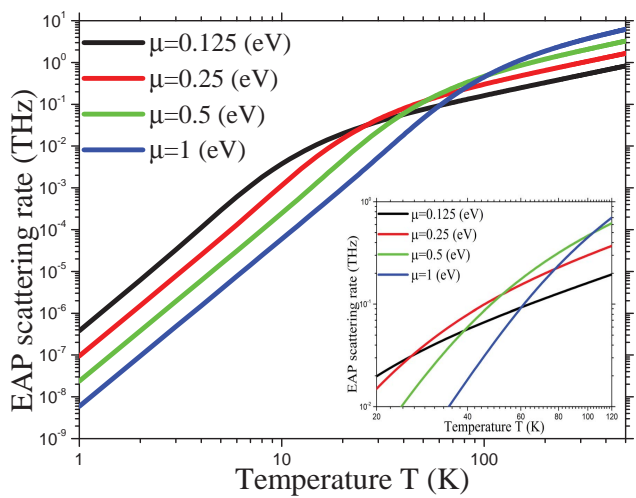}
\caption{\label{s8} Graphene's in-plane acoustic scattering rate as a function of temperature is calculated at $\mu$=0.125 eV (in black), $\mu$=0.25 eV (in red), $\mu$=0.5 eV (in green), and $\mu$=1 eV (in blue). The inset is an enlarged part for 20K $\le T \le$ 120K.}
\end{figure}

\begin{figure}[b]
\includegraphics[width=1.\linewidth]{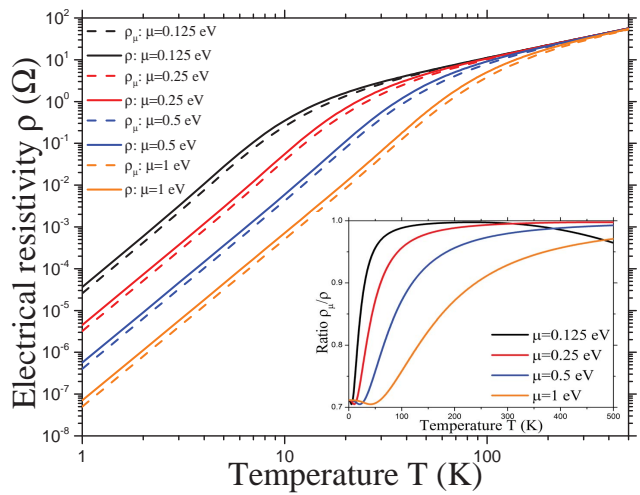}
\caption{\label{s5} The in-plane acoustic-phonon limited resistivity of graphene calculated by using the inelastic $\tau_{in}^{-1}(\mu)$ ($\rho_\mu$, the dash curves) and $\tau_{in}^{-1}(\epsilon_{sk})$ ($\rho$, the solid curves) scattering rates as a function of temperature at different chemical potentials. The inset shows their ratio $\rho_\mu/\rho$.}
\end{figure}

Due to this unusual behavior, a critical temperature of inelastic EAP scattering rates exists for a given $mu$ in graphene. Figure~\ref{s8} demonstrates the crossing of density-dependence of EAP scattering rates at a critical temperature $T_c$. It is found that the EAP scattering rate in graphene decreases with increasing carrier density when $T$ is lower than $T_c$, while beyond $T_c$ the scattering rate increases with carrier density. The value of $T_c$ depends on the range of $\mu$ considered  as shown in the inset. For $0.125$ eV $< \vert\mu\vert \le$ 0.25 eV, $T_c \approx 25K$, while  for $0.5$ eV $< \vert\mu\vert \le$ 1 eV, $T_c$ is as high as $\sim$ 100K.

Although the transport properties are dominated by the scattering rate evaluated at $\epsilon_{sk}=\mu$, Fig.~\ref{s5} shows the resistivity ($\rho_{\mu}$) of graphene calculated by using the inelastic EAP scattering rate at $\epsilon_{sk}=\mu$, $\tau_{in}^{-1}(\mu)$ by Eq.~(\ref{rate1}) (dash curves) and the resistivity ($\rho$) calculated by using energy-dependent $\tau_{in}^{-1}(\epsilon_{sk})$ by Eq.~(\ref{rate0}) (solid curves) at various chemical potentials can be quite different when $k_BT\ll \mu$. The ratio $\rho_\mu/\rho$ as a function of temperature is also shown in the inset. It is found that $\rho_\mu$ is a good approximation to $\rho$ only at high temperatures and low dopings, whereas $\rho_\mu$  reduces to around 70\% of $\rho$  at low temperatures and high dopings. These results agree with the analysis mentioned above about the doping and temperature effects.

\begin{figure}[t!]
\includegraphics[width=1.\linewidth]{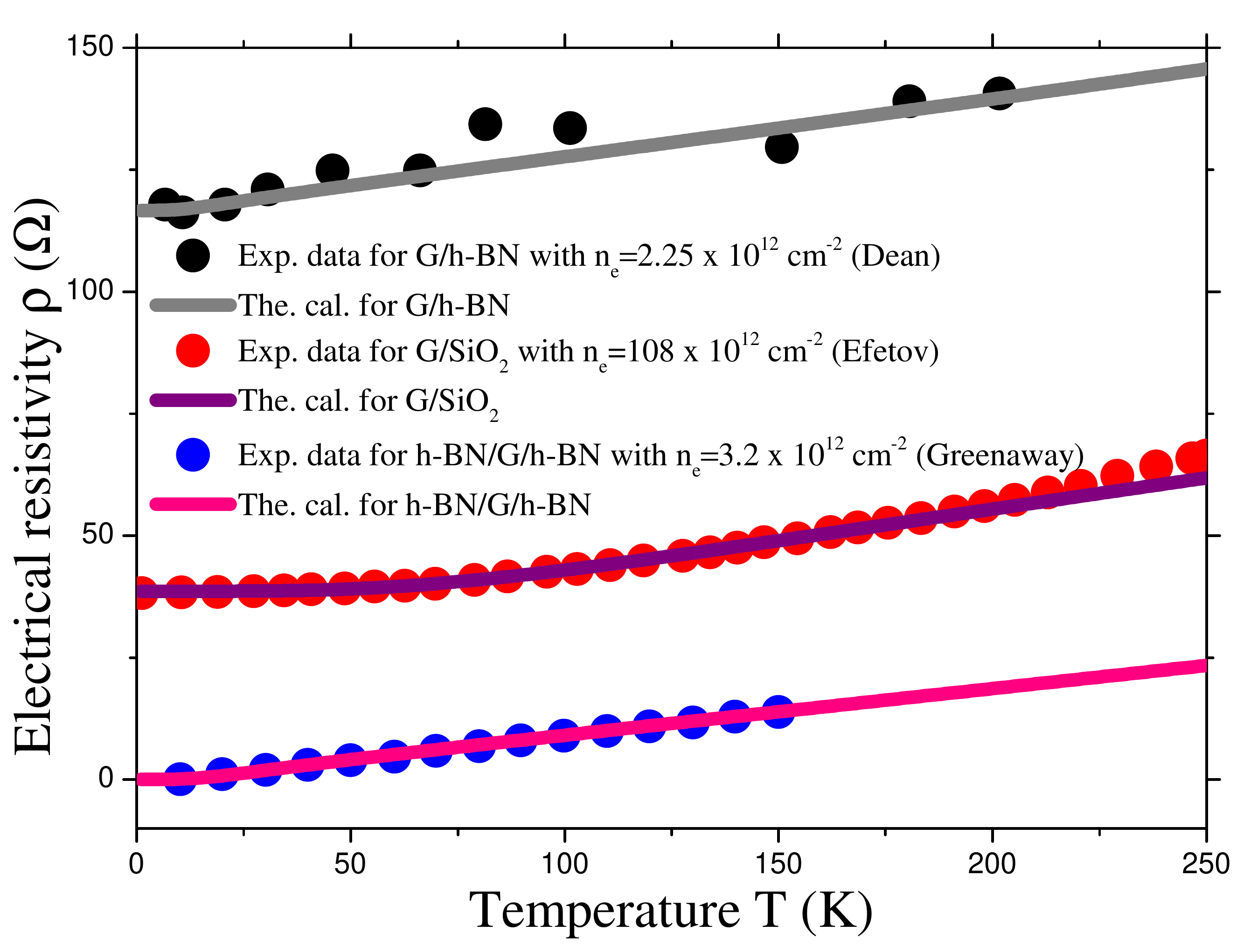}
\caption{\label{fig3}
Calculated electrical resistivity for graphene/h-BN with $n_e = 2.25 \times 10^{12}cm^{-2}$ (the gray curve) and data from Ref. \cite{Dean2010}, graphene/SiO$_2$ with $n_e=108 \times 10^{12}cm^{-2}$ (the purple curve) and data from Ref. \cite{Efetov2010}, and h-BN/graphene/h-BN with $n_e = 3.2 \times 10^{12}cm^{-2}$ (the pink curve) and data from Ref. \cite{Kumaravadivel2019,Greenaway2019}. Here we adopted $\beta=3$ for graphene/h-BN and graphene/SiO$_2$, while $\beta$ = 2.75 for h-BN/graphene/h-BN.}
\end{figure}

Finally, we compare theoretical predictions of our inelastic-scattering model to experimental data. Our calculated resistivity of graphene on different substrates by using $\rho = \sigma^{-1}$ with $\sigma$ given by Eq.~(\ref{tau})  are shown in Fig.~\ref{fig3} for graphene/h-BN with $n_e = 2.25 \times 10^{12}cm^{-2}$ (the gray line),  graphene/SiO$_2$ with $n_e=108 \times 10^{12}cm^{-2}$ (the purple line), and graphene sandwiched between h-BN with $n_e = 3.2 \times 10^{12}cm^{-2}$ (the pink line). For $T \lessapprox$ 200 K, $\rho(T, \mu)$ is predominantly due to acoustic-phonon scattering. Contributions from optical, zone-boundary phonons \cite{Park2014,Sohier2014}, and surface polar phonons for graphene/SiO$_2$ \cite{Li2010,You2019} should be taken into account when $T \gtrapprox$ 200 K. Our calculated results based on full inelastic scattering match experimental data for all three samples (with carrier densities up to 108 $\times 10^{12}\,cm^{-2}$) very well. Note that we have added a constant  scattering rate of 2.4 THz and 5.5 THz in fitting graphene/h-BN and graphene/SiO$_2$, respectively, to take into account effects of scattering mechanisms beyond EAP scattering.

Nonlinearity in $T$ was observed in $\rho(T, \mu)$ \cite{Tan2007,Morozov2008,Bolotin2008,Chen2008,DaSilva2010,Efetov2010,Park2014}, which was attributed to surface polar \cite{Chen2008,DaSilva2010}, flexural \cite{Morozov2008,Castro2010}, or optical and zone-boundary phonons \cite{Park2014}. Our equation Eq.~(\ref{rHT1}) suggests a  nonlinear-in-$T$ correction in high-$T$ regime when $\mu$ is comparable to $k_BT$, which should also be taken into account in such analyses.

\section{The validity of Matthiessen's rule}

\begin{figure}[b]
\includegraphics[width=1.\linewidth]{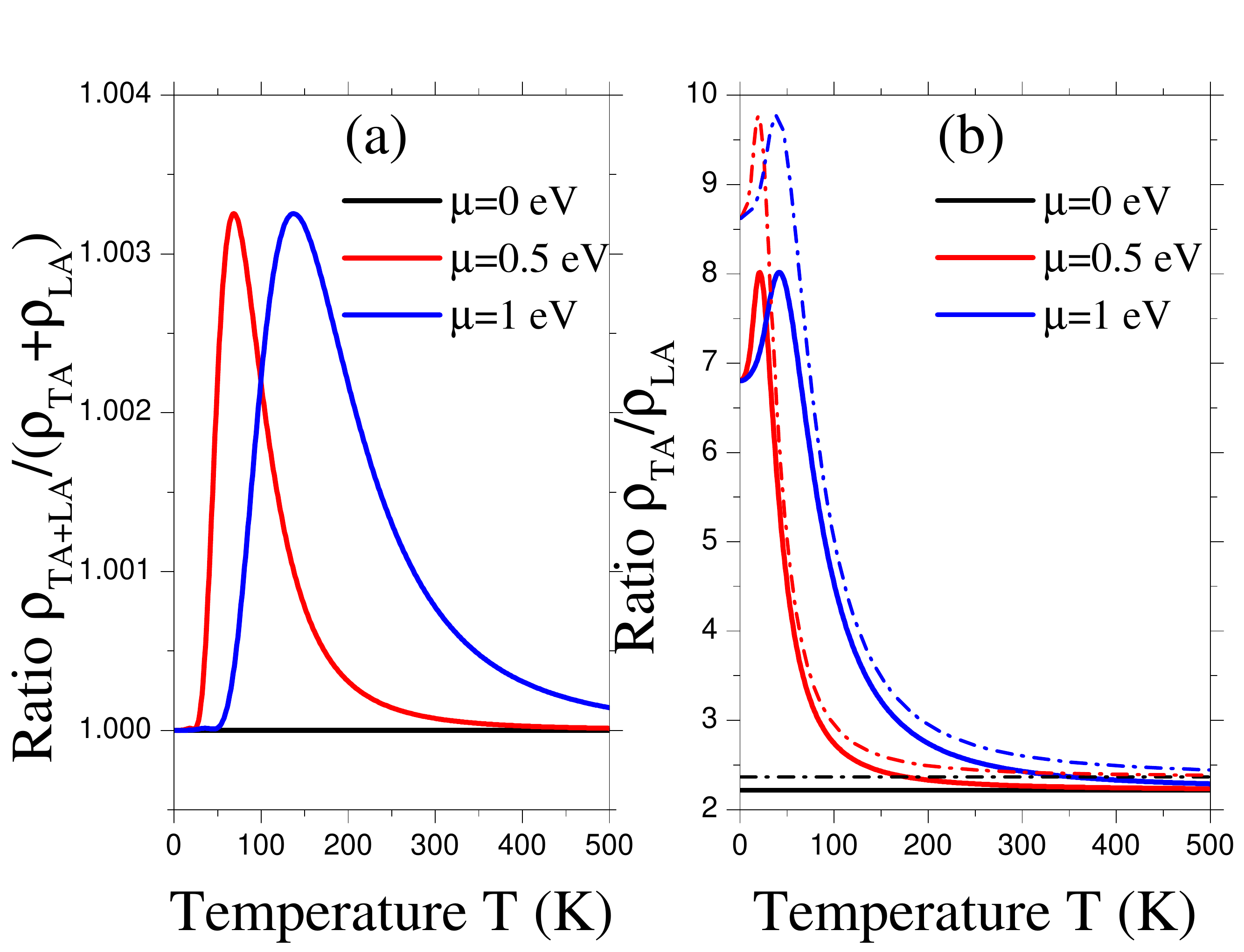}
\caption{\label{s6} The ratio (a) between the in-plane acoustic-phonon limited resistivity $\rho_{TA+LA}$ and the sum of seperate resistivities $\rho_{TA}+\rho_{LA}$ and (b) between $\rho_{TA}$ and $\rho_{LA}$ as a function of temperature at different chemical potentials. Note that the dash-dot curves in the panel (b) are the corresponding results when the contribution from the screened deformation potential is removed.}
\end{figure}

There is some debate in the literature about the validity of Matthiessen's rule \cite{Matthiessen1864} for the resistivity in 2D Dirac materials \cite{Hwang2008,Sohier2014}. Such a debate is difficult to resolve without an accurate assessment of the EAP scattering rates, especially at low temperatures and high dopings. Here, our calculations based on full inelastic equations can provide an answer to this question. Fig.~\ref{s6} shows (a) the ratio between the in-plane acoustic-phonon limited resistivity ($\rho_{TA+LA}$) and the sum of separate contributions to  the resistivity from LA and TA modes ($\rho_{TA}+\rho_{LA}$) and (b) the ratio between $\rho_{TA}$ and $\rho_{LA}$ as a function of temperature at various chemical potentials. Results in (a) show  a maximum error of $\sim$ 0.33\% in agreement with Ref. \cite{Sohier2014}, which validates Matthiessen's rule \cite{Matthiessen1864}. In (b), it is shown that $\rho_{TA}/\rho_{LA} >$ 2.2 which agrees with the previous results \cite{Park2014,Li2014,Sohier2014,Kumaravadivel2019,Greenaway2019}. It is worth mentioning that our inelastic equations can demonstrate the effects of doping and temperature and we find that the heavier graphene gets doped, the more TA phonons contribute versus LA phonons, especially at low temperatures. Note that the dash-dotted curves in (b) are the corresponding results when the contribution from the screened deformation potential is removed. Interestingly, for intrinsic graphene, we find $\rho_{TA}/\rho_{LA} \sim 2.4$ (which is also the ratio for doped graphene at high temperatures), in excellent agreement with the value of 2.5 reported in Ref. \cite{Park2014} by first-principles studies when the screened deformation potential is removed from consideration. Moreover, the contribution from the screened deformation potential at low temperatures ($T <$ 50 K) and finite dopings is quite significant. Finally, $\rho_{TA}/\rho_{LA}$ can be as high as $\sim$ 8 and $\sim$ 9.5 at low $T$'s and high $\mu$'s with and without considering the screened deformation potential, respectively.

\section{The validity of the conventional determination of the effective deformation potential in graphene}

\begin{figure}[b]
\includegraphics[width=1.\linewidth]{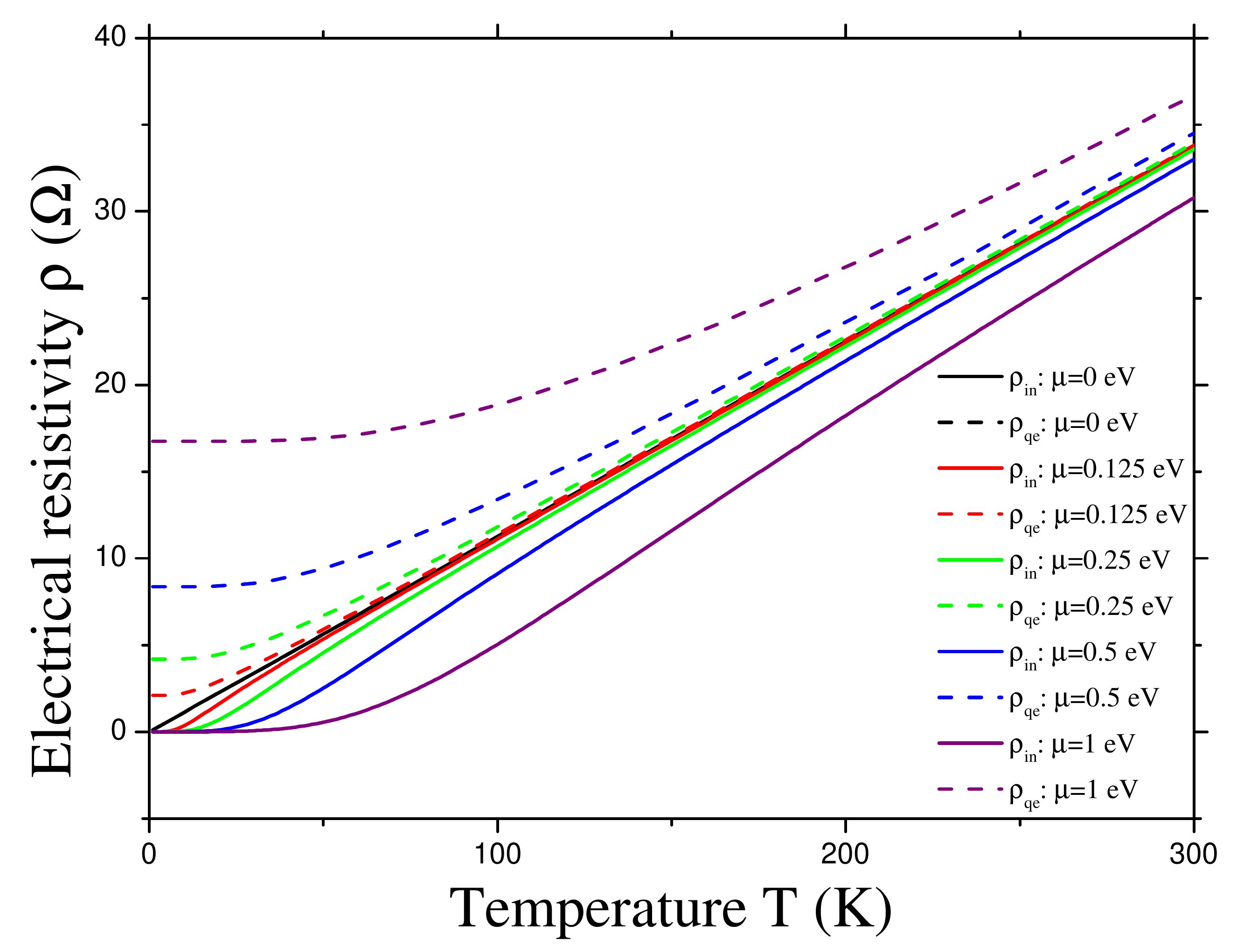}
\caption{\label{s7} Graphene's electrical resistivity is calculated from the inelastic (solid curves) and quasielastic (dash curves) scattering rate at $\mu$=0 eV (in black), $\mu$=0.125 eV (in red), $\mu$=0.25 eV (in green), $\mu$=0.5 eV (in blue), and $\mu$=1 eV (in purple). The same data as in Fig.~\ref{fig2} are replotted on the linear scale.}
\end{figure}

Conventionally, the effective deformation potential $J_a$ in graphene is determined from the slope of the linear part of the low-temperature resistivity $\rho(T, \mu)$ at a fixed carrier density (i.e. a fixed $\mu$) by using a Monte Carlo simulation \cite{Li2010} or by applying Eq.~(\ref{eq1}) \cite{Kaasbjerg2012,You2019}. However, because of the temperature and doping effects we have discussed above, the procedure of determining $J_a$ in Ref.~\cite{Li2010} is only valid for a given $\mu$, i.e. $J_a$ varies as a function of $\mu$ as shown in Fig.~\ref{s7}. And if Eq.~(\ref{eq1}) is used as in Refs.~\cite{Kaasbjerg2012,You2019}, $J_a$ is only valid for high temperatures and low chemical potentials. For the same input value of $J_a$ the slope of  $\rho(T, \mu)$ at low temperatures can deviate from $J_a$ significantly as implied in Fig.~\ref{s7}. This is the main reason why there have been so diverse values of the effective deformation potential $J_a$ in the literature, besides uncertainty in $v_{LA},\,v_{TA},\,\gamma_0\,(\mbox{or }v_F),\,\beta,\,\vert{E_1}\vert\,(\mbox{or } g_0/\epsilon(q))$. In fact, for intrinsic graphene such that Eq.~(\ref{eq1}) works perfectly because of $\hbar\omega_a/k_BT \ll 1$, the intrinsic effective deformation potential is given by
$J_a = \sqrt{ E_1^2 + 2B^2\left( 1 + v_{LA}^2/v_{TA}^2 \right)} \approx 16.7-18.2$ eV for $\beta = 2.75-3$, which is slightly higher than the in-plane value of the deformation potential in pristine graphite of 16.2 eV \cite{Ono1966}. No other universal effective deformation potentials exist because of the temperature and doping effects.

\section{Conclusion}

In conclusion, a full analytical consideration of  inelastic acoustic phonon scatterings in 2D Dirac materials for large range of temperature ($T$) and chemical potential ($\mu$) is presented to resolve several lingering issues on in-plane acoustic phonon scatterings in graphene. We shown analytically that the product of Bosonic and Fermionic distribution functions can be reduced to a simple expression, $\csch(x^p_a)$ after taking into account the momentum and energy conservations.  Acoustic phonon scattering rates versus $T$ for various doping  concentrations are investigated. Moreover, in both high-$T$ and low-$T$ limits, the well known quasielastic expressions of acoustic phonon scatterings are reproduced. We show that for heavily-doped graphene, the scattering rate in the high-$T$ limit is better expressed by Eq.~(\ref{rHT1}) (i.e. $\rho(T, \mu) \propto T[1 - \zeta_a\mu^2/3(k_BT)^2]$) than the linear-in-$T$ expression (\ref{eq1}) (i.e. $\mu$-independent linear-in-$T$ resistivity), which may account for the  nonlinearity in $T$ behavior observed in some experiments. In the low-$T$ limit, the $T^4$ dependence is revealed, although the prefactor derived here is different from the one reported previously \cite{Hwang2008,Efetov2010}. From our full inelastic expression, we can also extract an analytic semi-inelastic expression, which explains how the $T^4$ dependence gradually changes to linear-$T$ behavior as $T$ increases. It also explains why there are controversies in the low-$T$ behaviors as various $T^n$ behavior with $n=2,\,4,\,6$ \cite{Viljas2010,Lucas2016,Hwang2008,Mariani2008,Efetov2010,Mariani2010,Min2011,Sarma2011,Cooper2012} were reported in the low $T$ regime. This simple semi-inelastic expression can reproduce the full inelastic result for Dirac materials at any $T$ and $\mu$ of interest as long as $v_a/v_F \ll 1$ is satisfied. For intrinsic and lightly-doped graphene, the well-known behavior of $k_BT/\hbar\omega$ still holds for low $T$'s when $\vert\mu\vert \ll {v_F k_BT}/{2v_a} \approx 25 k_BT$; thus naming it the high-$T$ EAP scattering rate implying the scattering rate for only high $T$'s as elsewhere in the literature is not correct. Our results agree with previous first-principles studies and experimental data. Moreover, our analyses provide a more reliable way to determine the transition of $\rho(T, \mu)$ from $k_BT$ in the high $T$ regime to $(k_BT)^4$ in the low $T$ region, which allows a more meaningful determination of $\Theta_{BG}^a$ experimentally. In addition, the contributions from LA and TA modes at different temperatures and dopings are also analyzed in details, i.e. 2 $< \rho_{TA}/\rho_{LA} <$ 10; we also infer the analytical form of the acoustic gauge field \cite{Park2014} $\beta_A = B/\sqrt{2} = 3\beta\gamma_0/4\sqrt{2}$ and discuss the validity of Matthiessen's rule \cite{Matthiessen1864} and of the conventional determination of the effective deformation potential \cite{Li2010,Kaasbjerg2012,You2019}. Interestingly, contrary to the common perception, $\rho_\mu$ may contribute to total $\rho$ as low as $\sim$ 70\% at low $T$'s and, especially, high $\mu$'s; this weird behavior comes from decreasing of EAP scattering rate with increasing carrier density at low $T$'s. Finally, our studies pave a way for investigating scatterings between electrons and other fundamental excitations with linear dispersion relation in 2D Dirac material-based heterostructures such as bogolon-mediated electron scattering in graphene-based hybrid Bose-Fermi systems \cite{Sun2019}.

\begin{acknowledgments}
Work supported in part by Ministry of Science and Technology (MOST), Taiwan  under contract nos. 107-2112-M-001-032 and 108-2112-M-001-041.
\end{acknowledgments}

%\bibliography{nvk1ref}% Produces the bibliography via BibTeX.

\end{document}